\documentclass[useAMS,usenatbib]{mn2e}
\usepackage{latexsym,graphicx,natbib,amssymb}

\title[Velocity dispersion and M/L ratio of NGC 2419]
 {The velocity dispersion and mass-to-light ratio of the remote halo globular cluster NGC 2419}
      
\author[Baumgardt et al.]
{{\Large H. Baumgardt$^{1}$, P. C\^ot\'e$^2$, M.\ Hilker$^3$, M. Rejkuba$^3$, S. Mieske$^4$, S. G. Djorgovski$^5$ and Peter Stetson$^2$}\\
 $^1$ Argelander-Institut f\"ur Astronomie, Universit\"at Bonn, Auf dem H\"ugel 71, 53121 Bonn, Germany \\
 $^2$ National Research Council of Canada, Herzberg Institute of Astrophysics, 5071 West Saanich Road, Victoria, BC V9E 2E7, Canada \\
 $^3$ European Southern Observatory, Karl-Schwarzschild-Strasse 2, 85748 Garching bei M\"unchen, Germany \\
 $^4$ European Southern Observatory, Alonso de C\'ordova 3107, Vitacura, Casilla 19001, Santiago, Chile \\
 $^5$ Astronomy Department, California Institute of Technology, MC 105-24, Pasadena, CA 91125, USA
}

\date{Accepted ????. Received ?????; in original form ?????}

\pagerange{\pageref{firstpage}--\pageref{lastpage}} \pubyear{2008}

\begin{document}   

\maketitle

\label{firstpage}
 
\begin{abstract}
Precise radial velocity measurements from HIRES on the Keck I telescope are presented for 40 stars in the outer halo globular cluster
NGC~2419. These data are used to probe
the cluster's stellar mass function and search for the presence of dark matter in this cluster. NGC~2419 is one of the best
Galactic globular clusters for such a study due to its long relaxation time ($T_{r0} \approx 10^{10}$~yr) and large Galactocentric 
distance ($R_{GC} \approx 90$~kpc) --- properties
that make significant evolutionary changes in the low-mass end of the cluster mass function unlikely.
We find a mean cluster
velocity of $<\!\!v_r\!\!>=-20.3 \pm 0.7$ km/sec and an internal velocity dispersion of $\sigma = 4.14 \pm 0.48$ km/sec,
leading to a total mass of $(9.0 \pm 2.2) \cdot 10^5$ M$_\odot$ and a global mass-to-light ratio of $M/L_V = 2.05 \pm 0.50$ in solar units. This mass-to-light ratio is in good agreement with what one would expect 
for a pure stellar system following a standard mass function at the metallicity of NGC~2419. In addition, the 
mass-to-light ratio does not appear to rise towards the outer parts of the cluster. Our measurements therefore rule out the
presence of a dark matter halo with mass larger than $\sim 10^7$ M$_\odot$ inside the central 500 pc, which is
lower than what is found for the central dark matter densities of dSph galaxies. We also discuss 
the relevance of our measurements for alternative gravitational theories such as MOND, and for
possible formation scenarios of ultra-compact dwarf galaxies.
\end{abstract}


\section{Introduction}
\label{sec:intro}

An important unanswered question in modern astrophysics concerns the mechanism(s) by which stars and star clusters
form. This is especially true in the case of globular clusters, which, due to their large ages, are probes of 
the evolution and assembly history of galaxies in the early universe. 

Various theories have been proposed concerning the formation process of globular clusters. 
A number of authors proposed that globular clusters formed at the centers of dark matter halos (e.g.\ Peebles 1984, West 1993, Cen 2001,
Kravtsov \&  Gnedin 2005). Such a cosmologically-motivated formation scenario for globular clusters is indicated by the
extreme age of most clusters, their low metallicities, and their characteristic masses (which are close to the Jeans mass at recombination).
The fact that most clusters in the Milky Way contain no measurable amount of dark matter
is not in contradiction to these theories, since, over the course of a Hubble time, dark matter would get depleted
from the central regions of globular clusters due to the dynamical friction of stars against the much lighter dark matter
particles \citep{bm08}. As a result, dark matter tends to get pushed towards the outer cluster parts, where it is much more difficult to detect
and can be more easily stripped away by the tidal field of the Milky Way (e.g., Mashchenko \& Sills 2005).

A significant dark matter content might also be the explanation for the high mass-to-light ratios of ultra-compact dwarf galaxies (UCDs),
which have been discovered in spectroscopic surveys of nearby galaxy clusters since the late 1990s \citep{hetal99,detal00}. UCDs are
bright ($-11 \lesssim M_V \lesssim -13.5$~mag) and compact ($7 \lesssim r_h \lesssim 100$~pc) stellar systems that: (1) appear to obey 
a set of structural scaling relations distinct from those of globular clusters; and (2) have mass-to-light ratios that are,
on average, about twice as large than those of globular clusters of comparable metallicity, and somewhat larger than what one
would expect based on simple stellar evolution models assuming a standard stellar initial mass function \citep{hetal05,dhk08,metal08}.
One possible formation scenario for UCDs could be adiabatic gas infall into the center of a dwarf galaxy, which also funnels dark
matter into the center \citep{getal08}. Later tidal disruption of these galaxies would leave only the central star cluster behind. Since some of
the most massive Galactic globular clusters have masses and sizes only slightly smaller than UCDs 
(e.g. Rejkuba et al.\ 2007), it is conceivable that some of them 
may have formed in a similar way.

On the other hand, observations of interacting and starburst galaxies have shown that star clusters with masses comparable to globular clusters 
also form during major mergers between galaxies \citep{ws95,wetal99,getal01} in the local universe, which makes it possible that
globular clusters formed by similar processes in the early universe (e.g. Searle \& Zinn 1978, Ashman \& Zepf 1992). In 
such a scenario, globular cluster formation is driven mainly by gas-dynamical processes and the globular clusters should not contain 
significant amounts of dark matter.


Globular clusters in the outer halo of the Milky Way are ideal objects to test for the presence of dark matter and therefore examine the 
formation mechanism of globular clusters. Because the tidal field is weak in the remote halo, tidal stripping of the outer
cluster regions will be much less important than for globular clusters orbiting close to the Galactic center.
\citet{bm08} have shown that, due to dynamical friction,
dark matter gets depleted from the central regions of a star cluster on a timescale
\begin{equation}
\nonumber T_{\rm Fric} = 5.86 \left(\frac{M_{Tot}}{10^6 M_\odot}\right)^{1/2}  \left(\frac{r_h}{5 pc}\right)^{3/2} 
 \left(\frac{m}{M_\odot} \right)^{-1} \rm{Gyr} \,\, ,
\label{tfricgc}
\end{equation}
where $M_{Tot}$ is the total cluster mass, $r_h$ is the half-mass radius and $m$ is the average stellar mass in the cluster.
For clusters with $T_{\rm Fric} < T_{\rm Hubble}$, dark matter will be removed from the central parts within
a Hubble time.

\begin{figure}
\begin{center}
\includegraphics[width=8.5cm]{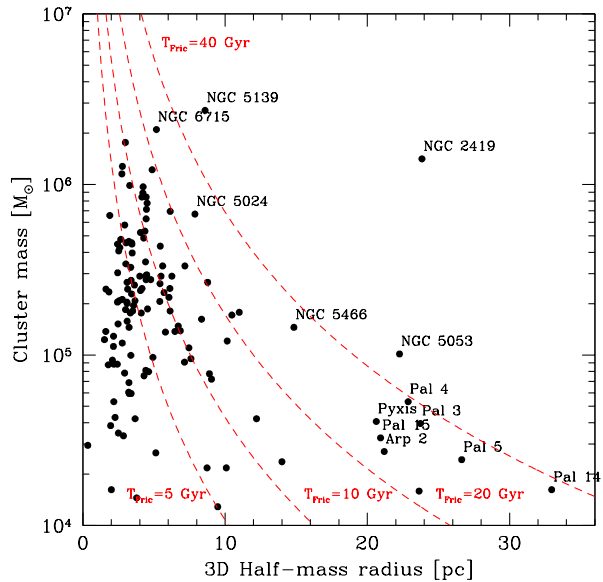}
\end{center}
\caption{Dynamical friction time, $T_{Fric}$, of stars against lighter dark matter particles for Galactic globular clusters. Most globular clusters 
have friction timescales of less than a Hubble time, meaning that dark matter would have been depleted from their centers if they formed
as a mix of dark matter and stars. Only a few extended clusters have friction times longer than a Hubble time and should therefore still retain
dark matter in their centers. With the longest friction timescale of all Galactic globular clusters, NGC~2419 is a promising target for a 
search for dark matter.}
\label{fig:tfric}
\end{figure}

Fig.~\ref{fig:tfric} shows the dynamical friction timescales of Galactic globular clusters. Here the cluster masses and half-mass radii, $r_h$, were calculated from
the absolute magnitudes, projected half-mass radii $r_{hp}$ and distances given by \citet{h96}, assuming 
$M/L=2.5$, which is appropriate for most metal-poor clusters with $[Fe/H]<-1.2$ and 
$r_h = 1.33 r_{hp}$, which is approximately correct for most King models.
It can be seen that
the majority of galactic globular clusters have friction timescales much less than a Hubble time. According to the results of 
\citet{bm08}, dark matter would exist in such clusters 
only in the outer parts, where it is difficult to detect due to low stellar densities and the influence of the Galactic tidal field
on the stellar velocities. Only a few clusters have friction timescales significantly longer than a Hubble time and are therefore good candidates
to look for primordial dark matter. Among the clusters with long dynamical friction timescales are Pal~3, Pal~4 and Pal~14, which are currently
being investigated by \citet{jetal09} in an effort to test alternative gravity theories like MOND \citep{m83a,m83b}. 

The Galactic globular cluster with the 
longest dynamical friction time, and therefore the smallest amount of dark matter depletion, is NGC~2419. With a Galactocentric distance of $R_{GC} \approx 90$ kpc,
NGC 2419 is also one of the most remote Galactic globular clusters, meaning that tidal stripping was likely to be least effective in this cluster.
Finally, NGC~2419, as one of the most massive Galactic globular clusters (see Fig.\ref{fig:tfric}), represents a possible Local Group analogue to a UCD. Its 
half-light radius of $r_{hp} = 47.5\arcsec \approx 19$~pc\footnote{At our adopted distance for NGC~2419, $(m-M)_0 = 19.60$~mag (Ripepi et al. 2007), 1\arcsec corresponds to 0.40~pc.}  (see \S\ref{sec:res-surf}) is also $\sim 6\times$ larger than is typical for globular 
clusters in the Milky Way and external galaxies (Jord\'an et~al. 2005), although still a factor of $\sim$ 2 smaller than the most extreme UCDs (see, e.g., Table~5 of Mieske et al. 2008).
For all of these reasons,  NGC~2419 is an excellent place to search for the presence of dark matter and thereby test different formation scenarios of 
globular clusters and, possibly, UCDs.

The only published velocity dispersion for NGC~2419 is that of \citet{oetal93} who, based on MMT radial velocity measurements with a median 
precision of 1.6 km/sec for 12 stars, found a mean radial velocity of $v_c = -20.0 \pm 0.9$ km/sec and an intrinsic velocity dispersion of 
$\sigma_c = 2.7 \pm 0.8$ km/sec. Fitting single-mass, isotropic King (1962) models to their radial velocities and the cluster surface brightness profile
yielded a mass-to-light ratio of $M/L_V \approx 0.7 \pm 0.4$~$M_{\odot} / L_{V,\odot}$ --- one of the smallest values measured for a Galactic 
globular cluster and much lower than the $M/L_V$s found for some UCDs.

In this paper, we report on radial velocity measurements for 44 candidate red giant branch (RGB) stars in the direction of NGC 2419 taken with
the High Resolution Echelle Spectrometer (HIRES; Vogt et al. 1994) on the Keck I telescope. The paper is organized as follows: In \S\ \ref{sec:obs} 
we describe the details of the observations 
and data reduction. In \S\ \ref{sec:results} we determine the total cluster luminosity, velocity dispersion and mass-to-light ratio 
of NGC 2419, and in \S\ \ref{sec:concl} we draw our conclusions.

\section{Observations and data reduction}
\label{sec:obs}

Candidate RGB stars in NGC~2419 were selected for observation with HIRES  from the
photographic study of Racine \& Harris (1975), as well as from unpublished CCD photometry of Peter Stetson. Fig.~\ref{fig:finder} presents 
a finding chart for the candidate RGB stars that were observed  for this program. 

HIRES spectra were obtained on four separate nights during two observing runs on the Keck I telescope. The dates of the two
runs were February [10-11] and March [9-11] 1999 (i.e., five nights in total). During each run, we limited the entrance aperture using the C1 decker
($0.866\times7.0\arcsec$) and binned the detector $1\times2$ (i.e., in the spatial direction) to reduce the read noise. The spectral
resolution for this instrumental configuration is $\cal R$ = 45000. A single readout amplifier was used (gain = 2.4 ADU$^{-1}$) with
the red collimator and a cross-disperser in first order. Thorium-Argon comparison lamp spectra were acquired frequently during
each night. In all, spectra were acquired for 44 different RGB
candidates during the two runs, with exposure times ranging from 180 to 600 sec. Approximately two dozen high-S/N
spectra for seven distinct IAU radial velocity standard stars were also obtained during the course of the program. 

Spectra for program objects and standard stars were reduced in an identical manner following the general procedures
described in C\^ot\'e et al. (1999, 2002). Briefly, the radial velocity of each RGB candidate was measured by cross-correlating 
its spectrum against that of a master template created during each run from the observations of IAU standard stars. In order to
minimize possible systematic effects, a master template for each observing run was derived from an
identical subsample of IAU standard stars. From each cross-correlation function, we measured both $v_r$, the heliocentric radial
velocity, and $R_{TD}$, the Tonry \& Davis (1979) estimator of the strength of the cross-correlation peak. 

The NGC~2419 observations discussed here
were taken as part of a broader program to study the internal dynamics of outer halo globular clusters, some results of which have
been reported elsewhere (i.e., C\^ot\'e et al. 2002, Jordi et al. 2009). During this program, which spanned seven Keck observing runs 
in 1998 and 1999 (13 nights in total), we obtained 53 distinct radial velocity measurements for 23 different RGB and subgiant stars belonging to 
Palomar 3, Palomar 4, Palomar 5, Palomar 14, and NGC 7492, in addition to NGC~2419. Using these repeat measurements and
following the procedures described in Vogt et al. (1995), we derived an empirical relationship between
our radial velocity uncertainties, $\epsilon(v_r)$, and the strength of cross correlation peak: $\epsilon(v_r) = \alpha / (1+R_{TD})$, 
where  $\alpha=9.0^{+2.4}_{-1.6}$ km/sec (90\% confidence limits).

Our measurements are summarized in Table~\ref{tabradvel}. The name of each program star is recorded in Column (1), with  
most identifications taken directly from Racine \& Harris (1975). The exception is for stars with IDs beginning with an
{\tt `S'} prefix, which denotes targets selected from our (unpublished) photometric catalogue. The second column of
this table gives the ID from {\citet{oetal93} if the star appeared in the earlier MMT study. Columns (3)-(9) record the position
in right ascension and declination, the radial distance from the 
cluster center, V magnitude and (B-V) colour from our catalogue, the number of independent HIRES radial velocity measurements,
and the adopted heliocentric radial velocity, $v_r$. For stars with multiple velocity measurements, the velocity reported in Column (9) refers to
the weighted mean of the individual measurements. We assumed that the cluster center is right ascension (J2000) 07 38 08.47 and
declination +38 52 56.8 in order to calculate radial distances of stars.

\begin{table*}
\caption[]{Radial velocities for candidate red giants in NGC 2419}
\begin{tabular}{lrc@{$\;\;$}c@{$\;\;$}cc@{$\;\;$}c@{$\;\;$}crccrcc}
\hline\hline
  ID  & 093 & \multicolumn{3}{c}{R.A.} & \multicolumn{3}{c}{Decl.}  & \multicolumn{1}{c}{R}     &   V   & (B-V) & $N_{v_r}$  &  $v_r$     &  Member? \\
  & & \multicolumn{3}{c}{(J2000.0)} & \multicolumn{3}{c}{(J2000.0)}  & (arcsec) & (mag) & (mag) & &  (km s$^{-1}$) & \\
 (1) & (2) & \multicolumn{3}{c}{(3)} & \multicolumn{3}{c}{(4)} &  \multicolumn{1}{c}{(5)}    &  (6)  &  (7)  &   (8)    &    (9)     &  (10) \\
\hline 
 III89  &III89& 07 & 38 & 09.79 & +38 & 50 & 45.4 & 132.3 &  17.25 &  1.42 &  1 &  -26.05$\pm$0.92 & Y \\
 S3     &.....& 07 & 38 & 01.92 & +38 & 53 & 15.4 &  78.7 &  17.26 &  1.44 &  1 &  -26.52$\pm$0.61 & Y \\
 S2     &.....& 07 & 38 & 09.90 & +38 & 52 & 00.7 &  58.5 &  17.27 &  1.55 &  1 &  -15.56$\pm$0.53 & Y \\
 S4     &.....& 07 & 38 & 07.71 & +38 & 53 & 15.7 &  20.9 &  17.27 &  1.51 &  1 &  -24.27$\pm$0.66 & Y \\
 S5     &.....& 07 & 38 & 10.07 & +38 & 52 & 54.3 &  18.9 &  17.29 &  1.48 &  1 &  -16.41$\pm$0.67 & Y \\
 S12    &.....& 07 & 38 & 06.00 & +38 & 53 & 37.5 &  49.9 &  17.40 &  1.39 &  1 &  -22.16$\pm$0.63 & Y \\
 S6     &.....& 07 & 38 & 06.89 & +38 & 53 & 33.5 &  41.1 &  17.41 &  1.28 &  1 &  -16.24$\pm$0.71 & Y \\
 S14    &.....& 07 & 38 & 11.58 & +38 & 53 & 05.8 &  37.4 &  17.42 &  1.32 &  1 &  -12.22$\pm$0.61 & Y \\
 S9     &.....& 07 & 38 & 10.18 & +38 & 53 & 09.6 &  23.7 &  17.42 &  1.24 &  1 &  -15.95$\pm$0.56 & Y \\
 34     &.....& 07 & 38 & 04.90 & +38 & 51 & 50.6 &  78.2 &  17.43 &  1.35 &  1 &  -20.96$\pm$1.28 & Y \\
 S10    &.....& 07 & 38 & 16.92 & +38 & 53 & 35.0 & 105.8 &  17.43 &  1.35 &  1 &  -20.75$\pm$0.58 & Y \\
 S17    &.....& 07 & 38 & 08.01 & +38 & 52 & 53.1 &   6.5 &  17.46 &  1.22 &  1 &  -10.12$\pm$0.81 & Y:\\
 S15    &.....& 07 & 38 & 09.69 & +38 & 52 & 40.9 &  21.4 &  17.48 &  1.43 &  1 &  -16.79$\pm$0.58 & Y \\
 S20    &.....& 07 & 38 & 07.60 & +38 & 53 & 34.1 &  38.7 &  17.50 &  1.38 &  1 &  -23.66$\pm$0.55 & Y \\
 S16    &.....& 07 & 38 & 14.21 & +38 & 52 & 36.0 &  70.2 &  17.51 &  1.36 &  1 &  -19.99$\pm$0.46 & Y \\
 S22    &C37  & 07 & 38 & 08.80 & +38 & 52 & 42.9 &  14.4 &  17.51 &  1.31 &  1 &  -33.98$\pm$0.64 & Y:\\
 S23    &C80  & 07 & 38 & 09.57 & +38 & 53 & 38.7 &  43.8 &  17.54 &  1.32 &  1 &  -22.81$\pm$0.58 & Y \\
 S26    &.....& 07 & 38 & 05.90 & +38 & 52 & 52.0 &  30.4 &  17.57 &  1.38 &  1 &  -27.17$\pm$0.53 & Y \\
 10     &10   & 07 & 38 & 16.92 & +38 & 53 & 35.0 & 105.8 &  17.61 &  1.28 &  2 &  -19.58$\pm$0.56 & Y \\
 38     &.....& 07 & 38 & 02.87 & +38 & 52 & 24.1 &  73.1 &  17.61 &  1.41 &  3 &  -23.63$\pm$0.40 & Y \\
 3      &.....& 07 & 38 & 03.48 & +38 & 54 & 12.1 &  95.2 &  17.63 &  1.34 &  1 &   20.76$\pm$0.79 & N \\
 S31    &.....& 07 & 38 & 08.64 & +38 & 53 & 00.0 &   3.8 &  17.64 &  1.19 &  1 &  -23.69$\pm$0.43 & Y \\
 S38    &38   & 07 & 38 & 07.67 & +38 & 52 & 45.5 &  14.7 &  17.75 &  1.18 &  1 &  -10.22$\pm$0.43 & Y:\\
 S41    &.....& 07 & 38 & 06.32 & +38 & 52 & 38.7 &  31.0 &  17.79 &  1.27 &  1 &  -29.27$\pm$0.44 & Y \\
 IV93   &.....& 07 & 37 & 56.45 & +38 & 51 & 22.8 & 169.0 &  17.80 &  1.29 &  1 &  -20.64$\pm$0.69 & Y \\
 II23   &.....& 07 & 38 & 09.58 & +38 & 56 & 30.0 & 213.6 &  17.95 &  1.14 &  1 &  -20.67$\pm$0.89 & Y \\
 4      &.....& 07 & 38 & 05.94 & +38 & 53 & 50.8 &  61.6 &  17.97 &  1.27 &  2 &  -23.49$\pm$0.83 & Y \\
 31     &.....& 07 & 38 & 11.15 & +38 & 51 & 53.4 &  70.7 &  18.07 &  1.17 &  1 &  -18.02$\pm$1.08 & Y \\
 I93    &.....& 07 & 38 & 01.00 & +38 & 54 & 57.8 & 149.1 &  18.08 &  1.54 &  1 &   70.61$\pm$0.99 & N \\
 II118  &.....& 07 & 38 & 16.93 & +38 & 54 & 25.2 & 132.5 &  18.09 &  1.15 &  1 &  -20.75$\pm$1.04 & Y \\
 I48    &.....& 07 & 37 & 58.06 & +38 & 55 & 22.2 & 189.5 &  18.13 &  1.12 &  2 &  -18.47$\pm$0.71 & Y \\
 I105   &.....& 07 & 37 & 54.13 & +38 & 54 & 32.7 & 192.9 &  18.16 &  1.16 &  1 &  -21.43$\pm$0.97 & Y \\
 III86  &.....& 07 & 38 & 09.54 & +38 & 50 & 31.7 & 145.6 &  18.23 &  1.19 &  1 &  -18.89$\pm$0.99 & Y \\
 11     &.....& 07 & 38 & 17.81 & +38 & 54 & 06.1 & 129.2 &  18.23 &  1.10 &  1 &   48.83$\pm$0.64 & N \\
 III37  &.....& 07 & 38 & 13.72 & +38 & 50 & 00.5 & 186.7 &  18.29 &  1.05 &  1 &  -21.30$\pm$1.20 & Y \\
 36     &.....& 07 & 38 & 03.29 & +38 & 51 & 57.2 &  84.9 &  18.31 &  1.23 &  1 &  -20.90$\pm$1.29 & Y \\
 35     &.....& 07 & 38 & 03.55 & +38 & 51 & 51.9 &  86.7 &  18.32 &  1.18 &  1 &  -18.45$\pm$0.98 & Y \\
 III107 &.....& 07 & 38 & 15.55 & +38 & 51 & 13.5 & 132.3 &  18.39 &  1.06 &  1 &  -17.43$\pm$0.99 & Y \\
 I143   &.....& 07 & 37 & 59.55 & +38 & 54 & 18.2 & 132.2 &  18.39 &  1.39 &  1 &   45.81$\pm$0.62 & N \\
 12     &.....& 07 & 38 & 18.63 & +38 & 53 & 52.0 & 130.8 &  18.60 &  1.02 &  1 &  -17.70$\pm$1.16 & Y \\
 1      &.....& 07 & 38 & 02.83 & +38 & 54 & 01.7 &  92.5 &  18.61 &  1.01 &  1 &  -21.63$\pm$1.31 & Y \\
 24     &.....& 07 & 38 & 18.18 & +38 & 52 & 15.9 & 120.5 &  18.70 &  0.98 &  1 &  -21.68$\pm$1.11 & Y \\
 II59   &.....& 07 & 38 & 13.97 & +38 & 55 & 26.7 & 163.1 &  18.80 &  0.98 &  1 &  -23.20$\pm$1.11 & Y \\
 7      &.....& 07 & 38 & 11.93 & +38 & 54 & 19.4 &  92.0 &  19.00 &  0.95 &  1 &  -22.40$\pm$1.26 & Y \\
\hline \hline
\end{tabular}
\label{tabradvel}
\end{table*}

In the left panel of Fig.~\ref{fig:cmd} we show a V,(B-V) colour-magnitude diagram (CMD) for stars within 250\arcsec of NGC~2419 --- extending from the tip of the
RGB to the lower subgiant branch --- that shows the location of our program stars in the CMD. For comparison, the solid curve
shows an isochrone from Dotter et al. (2008) having $T=12.3$ Gyr, and matching the chemical abundances of the cluster: e.g.,
[Fe/H] = -2.32 dex and [$\alpha$/Fe] = +0.2 dex (e.g., Shetrone, C\^ot\'e \& Sargent 2001). In making this comparison, we have 
adopted a reddening of $E(B-V) = 0.10$~mag --- intermediate between the values of 0.11~mag and ~0.08~mag reported by Harris et al.
(1997) and Ripepi et al. (2007) --- and a de-reddened distance modulus of $(m-M)_0 = 19.60$~mag  (Ripepi et al. 2007). Note that all of 
our program stars, and those from the previous MMT program, are located within $\sim 2$~mag of the RGB tip. 
The right panel of Fig.~\ref{fig:cmd} shows a magnified view of this part of the CMD. The dashed lines show a region of 
$\pm0.15$~mag width centered on the isochrone --- an interval chosen to roughly match the observed width of the RBG sequence
and encompass all of the probable radial velocity members. 
 
Since NGC~2419 is located at a Galactic latitude of $b\simeq 25^{\circ}$, it is important to consider the possibility of contamination by 
foreground disk stars. To gauge the extent of such contamination, we show in the right panel of Fig.~\ref{fig:cmd} the results of
one simulation using the Besan{\c c}on Galaxy model (Robin et al. 2003), for field stars in the range 0-100 kpc and located within
a $500\arcsec\times500\arcsec$ region in the direction of NGC~2419. This simulation yields a total of 10 stars having magnitudes
and colours that would place them within the region bounded by the dashed lines in the right panel of Fig.~\ref{fig:cmd}. However,
as we shall see in \S\ref{sec:res-vel}, only two stars in this simulation have radial velocities within $\pm3\sigma$ of the cluster mean. 
Since our radial velocity survey is by no means complete within this simulated region, we conclude that our final sample should have 
minimal foreground contamination with 1 interloper, {\it at most}, expected within $\pm3\sigma$ of the cluster's systemic velocity.
The final column of Table~\ref{tabradvel} gives our division of the sample into cluster members and foreground stars. In most 
cases, the assignments are unambiguous, although three stars (S17, S22 and S38) are more problematical. We shall return to this issue in \S\ref{sec:results}.

\begin{figure}
\begin{center}
\includegraphics[width=8.5cm]{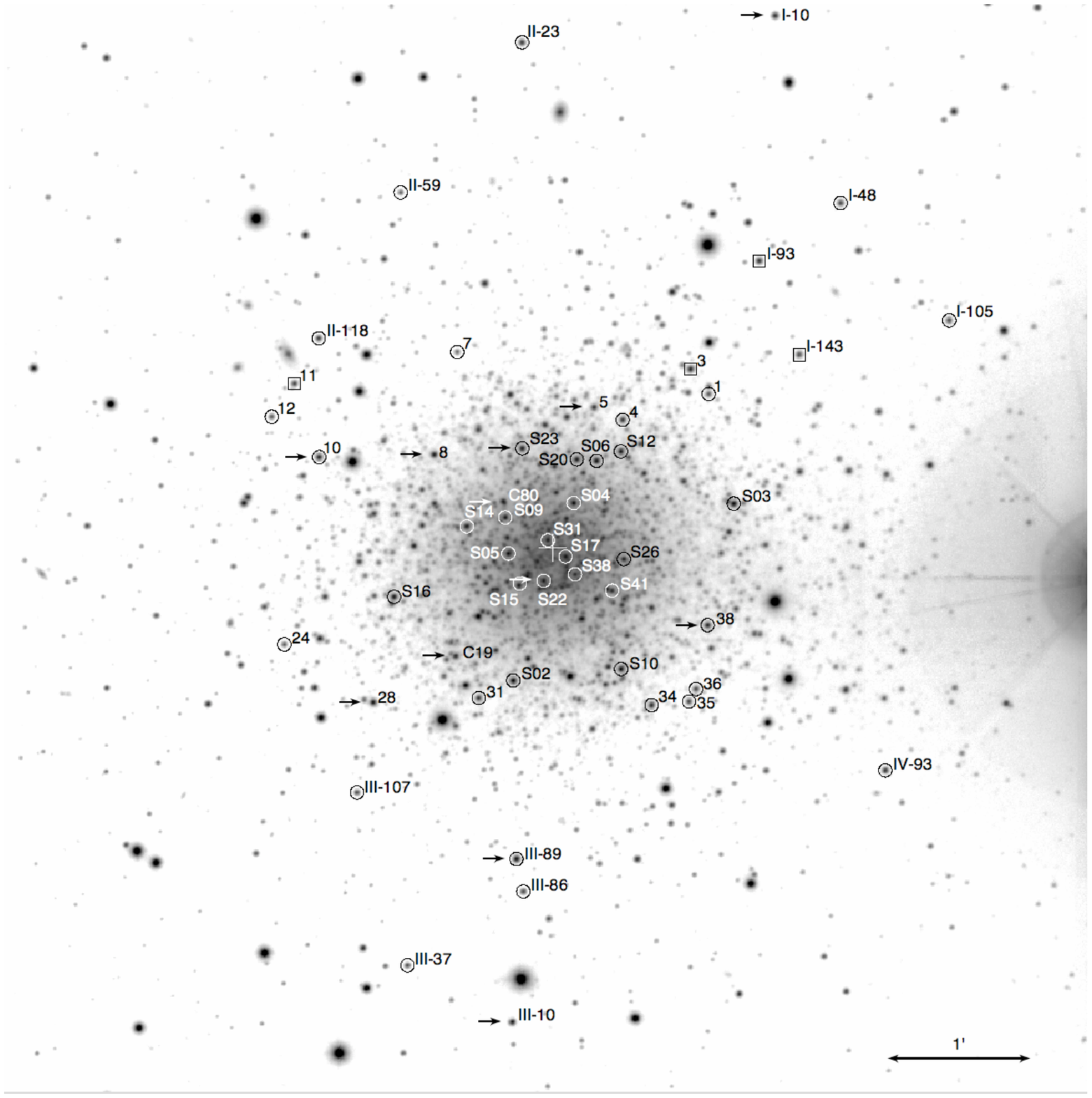}
\end{center}
\caption{V-band finding chart for NGC~2419 showing the location of candidate RGB stars with HIRES radial velocity measurements listed 
in Table~\ref{tabradvel}. Open circles and squares indicate cluster members and non-members, respectively. Arrows show the 
12 stars with MMT radial velocity measurements from \citet{oetal93}. The cross shows the adopted cluster center.
North is up and east is to the left in this image, which measures $7\farcm6\times7\farcm6$.}
\label{fig:finder}
\end{figure}

\begin{figure}
\begin{center}
\includegraphics[width=8.5cm]{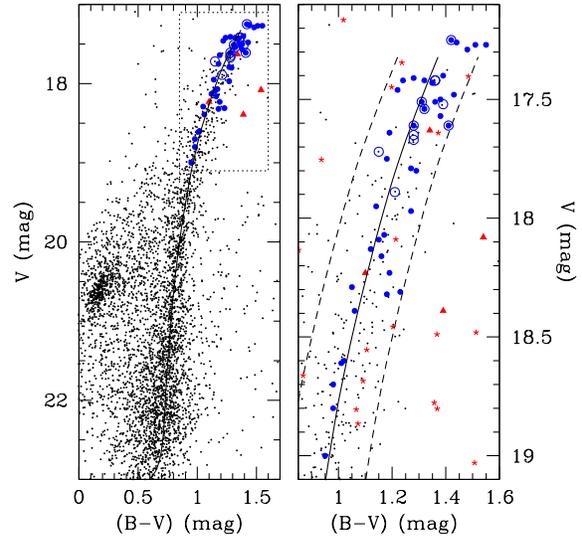}
\end{center}
\caption{Left panel: Colour-magnitude diagram for NGC~2419 from our own unpublished photometry, showing the location of our program stars. Filled blue circles 
show the 40 probable members according to our HIRES radial velocities. Open circles show the 12 stars with radial velocities measurements from 
Olszewski et al. (1993); there are five stars in common between the two studies. Red triangles show the four obvious non-members stars from
Table~\ref{tabradvel}. The smooth curve shows an 12.3 Gyr isochrone (Dotter et al. 2008) with [Fe/H] = -2.32 and [$\alpha$/Fe] = +0.2 dex, shifted
 by $E(B-V) = 0.10$~mag and  $(m-M)_V = (m-M)_0 + A_{\rm V}$ = 19.91~mag to match the RGB.
Right panel: Magnified view of the dotted region from the previous panel. The dashed curves show a $\pm$0.15~mag region centered on the isochrone. 
The red asterisks illustrate one simulation of the expected contamination in region of the colour-magnitude diagram from Galactic foreground stars in the direction of 
NGC~2419 using the Besan{\c c}on Galaxy model (Robin et al. 2003). See text for details.
}
\label{fig:cmd}
\end{figure}

\section{Results}
\label{sec:results}

\subsection{Surface density profile}
\label{sec:res-surf}

\begin{figure}
\begin{center}
\includegraphics[width=8.5cm]{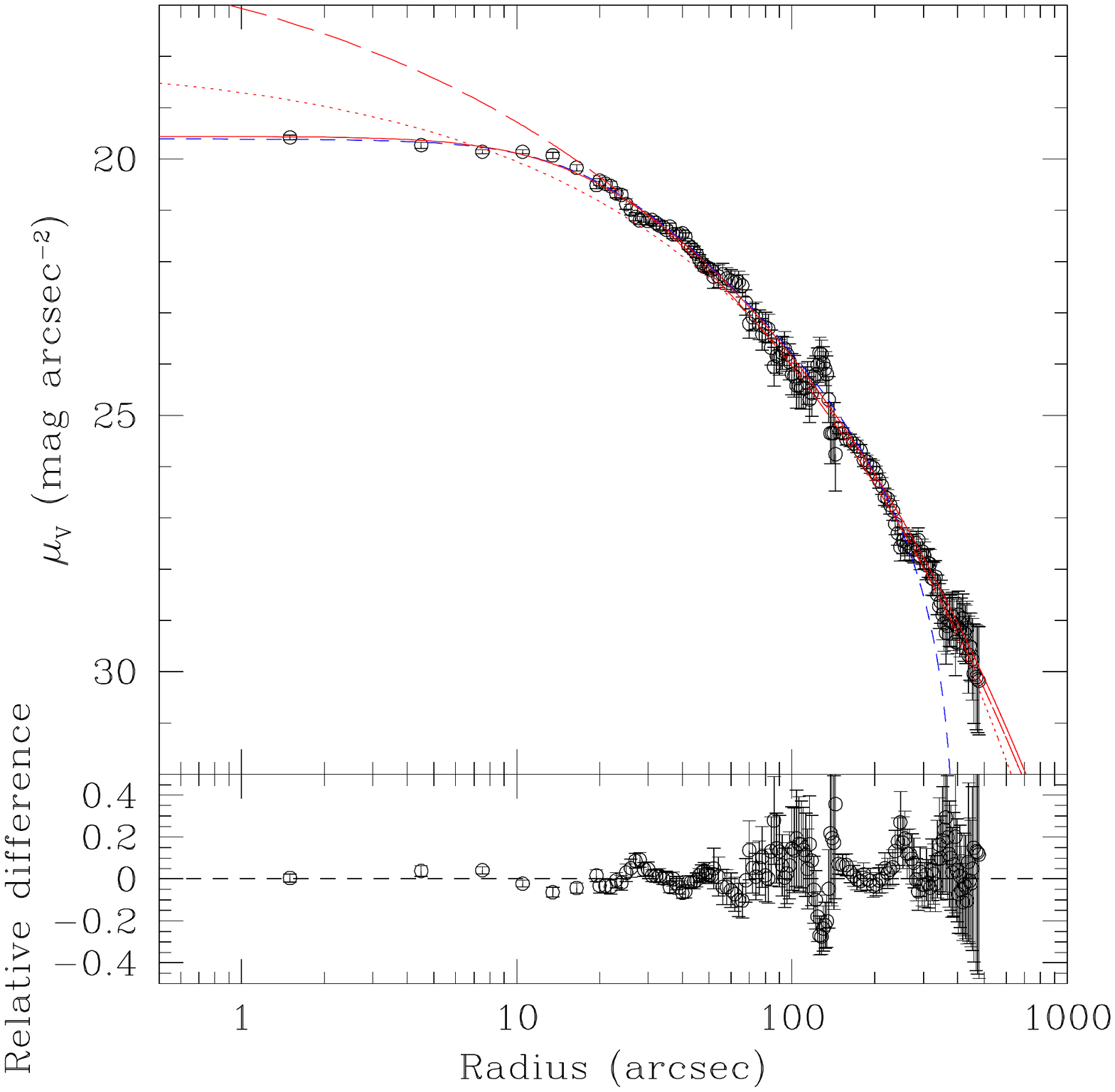}
\end{center}
\caption{Upper panel: V-band surface brightness profile of NGC~2419 as determined by Bellazzini (2007). The blue, short-dashed curve shows the
best fitting single-mass, isotropic King (1962) model. This model cannot fit the ``excess" light at large radii. The dotted and long-dashed red curves are
fits of two S\'ersic profiles that fit the outer parts of the cluster well but not the central region. The solid curve is a S\'ersic profile
with an added core of size $r_c =14\arcsec$. It provides a good match to the density profile and is within the reported
observational uncertainties at most radii (bottom panel).}
\label{fig:sdens}
\end{figure}

The surface brightness profile of NGC~2419 has most recently been determined by Bellazzini (2007) using
surface photometry and star counts from HST/ACS and SDSS. In this
paper we use his data for our analysis. Fig.\ \ref{fig:sdens} shows the V-band surface brightness profile of NGC 2419 as determined 
by Bellazzini (2007) and compares it with several different model profiles. Bellazzini (2007) found that the best-fitting King (1962) 
profile has a central core radius of $r_c=19.2^{\prime\prime}$ and a concentration index of $c=\log(r_t/r_c)=1.35$. We show this profile by the blue
dashed curve in Fig.\ \ref{fig:sdens}.  As can be seen, a King profile provides a satisfactory fit to the observed surface density
profile out to $\approx 300^{\prime\prime}$. Beyond this point, however, the observed brightness profile declines more slowly than the fitted King profile. Such an excess
could be the signature of extra-tidal stars (e.g., Leon et al. 2000). However, at the distance of NGC2419, 
300$\arcsec$ corresponds to a physical distance of about 120~pc. The expected 
tidal radius of NGC 2419 at its current distance 
is $r_t \sim$ 750 pc: i.e., more than 5 times as large, if we assume a circular cluster orbit and a logarithmic potential 
with circular velocity $v_c=200$ km/sec for the Milky Way and a cluster mass of $10^6$ M$_\odot$.
In addition, tidal radii often correspond to breaks in the surface density distribution but no such break is visible in Fig.~\ref{fig:sdens}.
Hence, unless the cluster orbit is highly eccentric ($\epsilon > 0.9$), the outermost stars are most likely bound to NGC~2419 so that the tidal 
radius has not yet been reached by the
observations, something which was also noted by \citet{retal07}.

We have therefore also tried to fit the surface brightness profile with S\'ersic (1968) profiles, which are known to provide accurate representations of the
surface brightness profiles of most early-type galaxies, including UCDs, over nearly their entire profiles 
(e.g., Graham \& Guzm\'an 2003, Ferrarese et al. 2006, C\^ot\'e et al. 2007, Evstigneeva et al. 2007, Hilker et al. 2007). 
We find that S\'ersic models that best fit the data over the whole
range (red dotted curve) and those fitted to the data for $r>20\arcsec$ both give good fits in the outer profiles but significantly overpredict
the central surface brightness. In order to obtain a model that fits the surface brightness profile over the whole radial range, we add a constant
density core to the S\'ersic profile fitted to the data for $r>20$\arcsec:
\begin{equation}
 \Sigma(r) = \Sigma_0 \; \exp(-(1.9992 n-0.3271) \cdot \left(r'/r_{e}\right)^{1/n}-1.0)
\end{equation} 
where $r'=\sqrt{r^2+r^2_c}$. A $\chi^2$ minimization gives 14$^{\prime\prime}$ as best fitting value for $r_c$. The solid red curve in 
Fig.\ref{fig:sdens} shows the surface 
brightness profile of the outer S\'ersic profile once
a constant density core with $r_c=14''$ has been added. With this model we obtain a very good fit to the overall surface
brightness profile. The lower panel shows the residuals between our fitted model and the observed data. At most radii, the differences between 
the observed surface density and our model are within the observational uncertainties.

Table~\ref{tab:surf} summarises our results. For a King model fit, we find that the best-fitting 
profile has $r_c=19^{\prime\prime}$ and $c=1.39$, which agrees reasonably well with 
Bellazzini (2007). The other rows list the parameters of our best-fitting S\'ersic profiles.
The integrated $V$ magnitudes are calculated separately for
each surface density profile. They are, in general, somewhat smaller than the values by \citet{b07}, who found 
V=$10.47 \pm 0.07$~mag. Because the King profile underpredicts the light in the outer parts while 
the two S\'ersic profiles significantly overpredict the light in the inner parts, we consider the values from the cored S\'ersic profile
to be most reliable and we will use them in the remainder of this paper. For the cored S\'ersic profile, we obtain a total luminosity
of V=$10.57$~mag and a projected half-light radius of $r_{hp}=47.5^{\prime\prime}$. The latter value agrees quite well with
the half-light radius found by \citet{tkd95}, $r_{hp}=45.7^{\prime\prime}$. With a distance modulus of $(m-M)_0=19.60 \pm 0.05$~mag
and a reddening of E(B-V)=0.08~mag \citep{retal07}, these values lead to a total absolute magnitude and projected half-mass radius of 
NGC~2419 of $M_V=-9.28$~mag and $r_{hp}=19.2$ pc respectively.
\begin{table}
\caption[]{Photometric parameters of best-fitting surface density profiles for NGC 2419}
\begin{tabular}{lccccc}
\hline \hline
 & c & $r_c$ & $r_{hp}$ & $\mu_{0}$ & $V_{tot}$ \\
 &  & [$^{\prime\prime}$] & [$^{\prime\prime}$] & [mag/sq.$^{\prime\prime}$] & (mag) \\
\hline
King (1962) & 1.39  & 19.0 & 44.7 &  19.61 & 10.65  \\
\hline \hline 
 & n & $r_e$ & $r_{hp}$ & $\mu_e$ & $V_{tot}$ \\
 &  & [$^{\prime\prime}$] & [$^{\prime\prime}$] & [mag/sq.$^{\prime\prime}$] & (mag) \\
\hline
S\'ersic (all r)    & 2.15  & 49.8 & 49.8 &  22.32 & 10.77  \\
S\'ersic ($r>20$\arcsec) & 3.01  & 33.2 & 33.2 &  21.31 & 10.48  \\
S\'ersic-core       & 3.01  & 33.2 & 47.5 &  21.23 & 10.57  \\
\hline \hline
\end{tabular}
\label{tab:surf}
\end{table}

\subsection{Velocity dispersion of NGC~2419}
\label{sec:res-vel}

Table~\ref{tabradvel} lists 44 stars in the direction of NGC~2419 with radial velocities measured in the course
of this study. Rejecting the four stars with the most discrepant velocities and consequently identified as 
non-members in Table~\ref{tabradvel} (e.g., S3, I93, 11 and I143) leaves us with a sample of 
40 stars having radial velocities in the range $-35 \le v_r \le -5$ km/sec. Although this sample shows
a clear peak at $\sim -20$ km/sec corresponding to the cluster systemic velocity, there are three 
stars that differ by $\sim$ 10 km/sec or more from the apparent cluster mean. The upper panel of
Fig.~\ref{fig:veldist}, which plots radial velocity against distance from the cluster center,
reveals all three to be among the centrally concentrated stars (i.e., with radii of $r \lesssim 15\arcsec$) and
thus unlikely to be interlopers. Nevertheless, we now pause to consider 
the extent to which our sample could be contaminated by foreground disk stars. 

As noted in \S\ref{sec:obs}, one simulation of the expected Galactic foreground using the Besan{\c c}on 
model (Robin et al. 2003) suggests we might expect {\it a total} of $\sim$ 10 interloping field stars: (1) 
within a $500\arcsec\times500\arcsec$ region along this line of sight; and (2) falling along the cluster RGB
shown in the right panel of Fig.~\ref{fig:cmd}. 
In the lower panel of Fig.~\ref{fig:veldist} we compare the observed velocity distribution of radial velocities
(open blue histogram) with that from the simulation (filled red histogram). The foreground stars plotted 
here represent all stars in this velocity range with magnitudes and colours that place them within the RGB 
locus shown in Fig.~\ref{fig:cmd}. Because our sample of 44 RGB candidates constitutes only a small 
fraction ($\approx 13\%$) of the full sample of stars in this region of the CMD and lying within $\sim$ 250\arcsec of the 
cluster core, we conclude that we would expect, at most, 1 interloper in our final sample of 40 cluster members.

We now proceed by calculating the average radial velocity and velocity dispersion of NGC 2419 using the method of 
Pryor \& Meylan (1993). We first assume that each velocity, $v_i$, is drawn from a normal distribution,
\begin{equation}
 f(v_i) = \frac{1}{\sqrt{2 \pi (\sigma_c^2 + \sigma^2_{e, i})}} exp \left( -\frac{(v_i-v_c)^2}{2(\sigma_c^2+\sigma_{e,i}^2)} \right) \; ,
\end{equation}
where $v_c$ and $\sigma_c$ are the cluster's average radial velocity and intrinsic velocity dispersion, and $\sigma_{e,i}$ is the 
individual velocity
error for each star. Calculating the likelihood function, $L$, for all stars and taking the partial derivatives of its logarithm $l=\log L$ 
with respect to $v_c$ and $\sigma_c$ leads to the following set of equations:
\begin{eqnarray}
\sum_{i=1}^N \frac{v_i}{(\sigma_c^2+\sigma^2_{e, i})} - v_c \sum_{i=1}^N \frac{1}{(\sigma_c^2+\sigma^2_{e, i})} & = & 0 \\
\sum_{i=1}^N \frac{(v_i-v_c)^2}{(\sigma_c^2+\sigma^2_{e, i})^2} - \sum_{i=1}^N \frac{1}{(\sigma_c^2+\sigma^2_{e, i})} & = & 0 
\end{eqnarray}
The above equations can be solved analytically to obtain a first guess value for $v_c$ and $\sigma_c$ if zero measurement errors 
$\sigma_{e, i}$ for all stars are assumed. 
The full solution can then be obtained iteratively, starting with the solution for zero measurement errors as first estimate. The errors
of $v_c$ and $\sigma_c$ are obtained from the information matrix, $I$, by
\begin{eqnarray}
 \sigma_v & = & I_{22}/(I_{11} I_{22}-I_{12}^2) \\
 \sigma_\sigma & = & I_{11}/(I_{11} I_{22}-I_{12}^2) 
\end{eqnarray}
where the components $I_{ij}$ of the information matrix are calculated as given in Pryor \& Meylan (1993).

Using the 40 probable members from Table~\ref{tabradvel}, we obtain a 
mean cluster velocity of $v_c=-20.63 \pm 0.74$ km/sec and an intrinsic velocity dispersion of 
$\sigma_c=4.61 \pm 0.53$ km/sec.
This mean velocity agrees very well with the one obtained by \citet{oetal93}. Our velocity dispersion is,
however, significantly higher than the value of $2.7 \pm 0.8$ km/sec determined by \citet{oetal93} and 
agrees with their value only at the 2$\sigma$ level. The explanation for this difference appears to lie in the radial
distribution of the stars in the two samples: Fig~\ref{fig:veldist} reveals the \citet{oetal93} sample to contain
only three stars in the central $r \sim 1\arcmin$, where we observe a significant rise in the cluster velocity dispersion
based on a larger sample of 17 stars in this region. In the outer parts of the cluster, our measured velocity dispersion (see below) is in 
very good agreement with the previous estimate.

\begin{figure}
\begin{center}
\includegraphics[width=8.5cm]{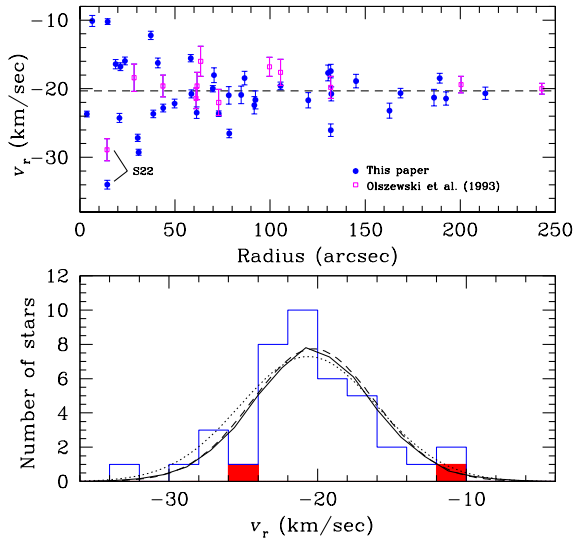}
\end{center}
\caption{Upper panel: Heliocentric radial velocity for NGC~2419 probable members plotted as a function of distance from the cluster center. 
The dashed line shows our best estimate for the cluster systemic velocity, $\langle v_r\rangle \approx -20.3$ km/sec. There is a clear rise in velocity
dispersion toward the cluster center. Note the discrepant measurements for star S22, which is the largest outlier in both our sample and 
that of \citet{oetal93}.
Lower panel: Radial velocity distribution of the measured stars in NGC 2419. Dotted and dashed lines show Gaussian fits to this distribution
obtained by either using all stars from Table~1 (dotted line), or by excluding star S22 --- which has the most discrepant velocity ---
from the sample (dashed line). The filled red histogram shows the expected distribution, in this velocity interval, for foreground stars 
within a $500\arcsec\times500\arcsec$ region along this line of sight and falling along the RGB in the cluster CMD (see the 
right panel of Fig.\ref{fig:cmd}). The solid line shows the distribution of stars from our model cluster fitted to the data without
star S22. It follows very closely the corresponding Gaussian distribution (dashed line).
}
\label{fig:veldist}
\end{figure}

The lower panel of Fig.\ \ref{fig:veldist} shows the distribution of radial velocities of the 40 candidate members in the
velocity range -36 to -4 km/sec (open histogram). The dotted curve shows a gaussian fit to
this distribution using the above values for $v_c$ and $\sigma_c$.
As can be seen, the observed radial velocity distribution is well approximated by this Gaussian except, as noted above, that the number 
of outliers in the wings of the distribution is somewhat higher than predicted: i.e., stars S17 and S38 are more 
than 2.2$\sigma$ away from the 
mean while star S22 is almost 3$\sigma$ away from the mean. For a Gaussian distribution and a sample of 40 stars, one would 
expect to find a star that is more than 3$\sigma$ 
away from the mean in only 10\% of all cases. Also, the chance to have three or more stars more than 2.2$\sigma$ away from the mean
is only 10\%, while the chance to have two such stars is twice has high. This makes it possible that at least one star, most
likely star S22, is either a cluster nonmember or a radial velocity variable. 

Since star S22 is located in the cluster center and 
has a photometry that places it squarely on the RGB (see the right panel of Fig.~\ref{fig:cmd}), it could be a 
binary system, in which case it should be excluded from the analysis. The absolute radial velocity
difference between star S22 and the cluster mean ($\sim$15 km/sec) is also so large that star S22 would be nearly unbound
if the radial velocity difference is due to a different orbital velocity. Star S22 is also one of the five stars 
in common with \citet{oetal93}, and has the most discrepant velocity in both samples. Moreover, 
two independent measurements different by 5.1 km/sec, or three times the quadrature sum of the individual uncertainties.
These facts suggest that star S22 may be a cluster binary or, perhaps more likely, an RGB star that shows a velocity ``jitter"
that has previously been found for some
globular cluster stars on the upper RGB (e.g., Gunn \& Griffin 1979, Mayor et al. 1984). 
If we omit star S22, we obtain a mean cluster velocity and an intrinsic velocity dispersion of:
\begin{equation}
\begin{array}{rrrl}
v_c & = & -20.28 \pm 0.68 & {\rm km/sec} \\
\sigma_c & = & 4.14 \pm 0.48 & {\rm km/sec}. \\
\end{array}
\end{equation}
These values are within the errorbars of the values that we obtain when using all stars and we will use them throughout the
rest of this paper. 

We note that the three stars with the most discrepant
radial velocities are also among the four most central stars. While this could simply be a statistical effect, it could also point to 
either a high binary fraction in the core or the broadening of the velocity distribution due to additional unseen mass,
possible if the cluster is mass segregated or contains an intermediate-mass black hole
\citep{betal05}. Multi-epoch observations of the stars in our sample and radial velocities of additional core stars would 
be required to test these interesting possibilities. 

\subsection{Cluster Rotation}
\label{sec:res-rot}

Our sample of 40 member stars also allows us to check for a possible rotation of NGC~2419. Fig.\ \ref{fig:rotvel} shows the radial velocities of 
stars as a function of position angle (PA). Here the position angle is measured from north (PA=$0^\circ$) towards east (PA=$+90^\circ$). There
is a clear dependence of the average velocity from the position angle visible in the data. This is confirmed by fitting a sinusoidal curve
to the radial velocity data. Allowing for a variable rotation angle, we obtain a rotation amplitude of $3.26 \pm 0.85$ km/sec around an axis 
with position angle of PA $=40.9 \pm 17.8$ degree. Our data therefore implies cluster rotation at more than the 3$\sigma$ level. The root mean square scatter
around the rotation curve is 4.00 km/sec. NGC~2419 is therefore partly rotationally supported and partly supported by random stellar motions.
\citet{b07} found that NGC~2419 is slightly elliptical with average ellipticity of $\epsilon=0.19 \pm 0.15$ and position angle of 
PA $= +105^\circ \pm 28^\circ$. Both the relatively small amount of ellipticity and the average angle of ellipticity, which is 
within the errorbars $90^\circ$ away from the rotation axis that we find, supports our finding that NGC~2419 is rotating.

\begin{figure}
\begin{center}
\includegraphics[width=8.5cm]{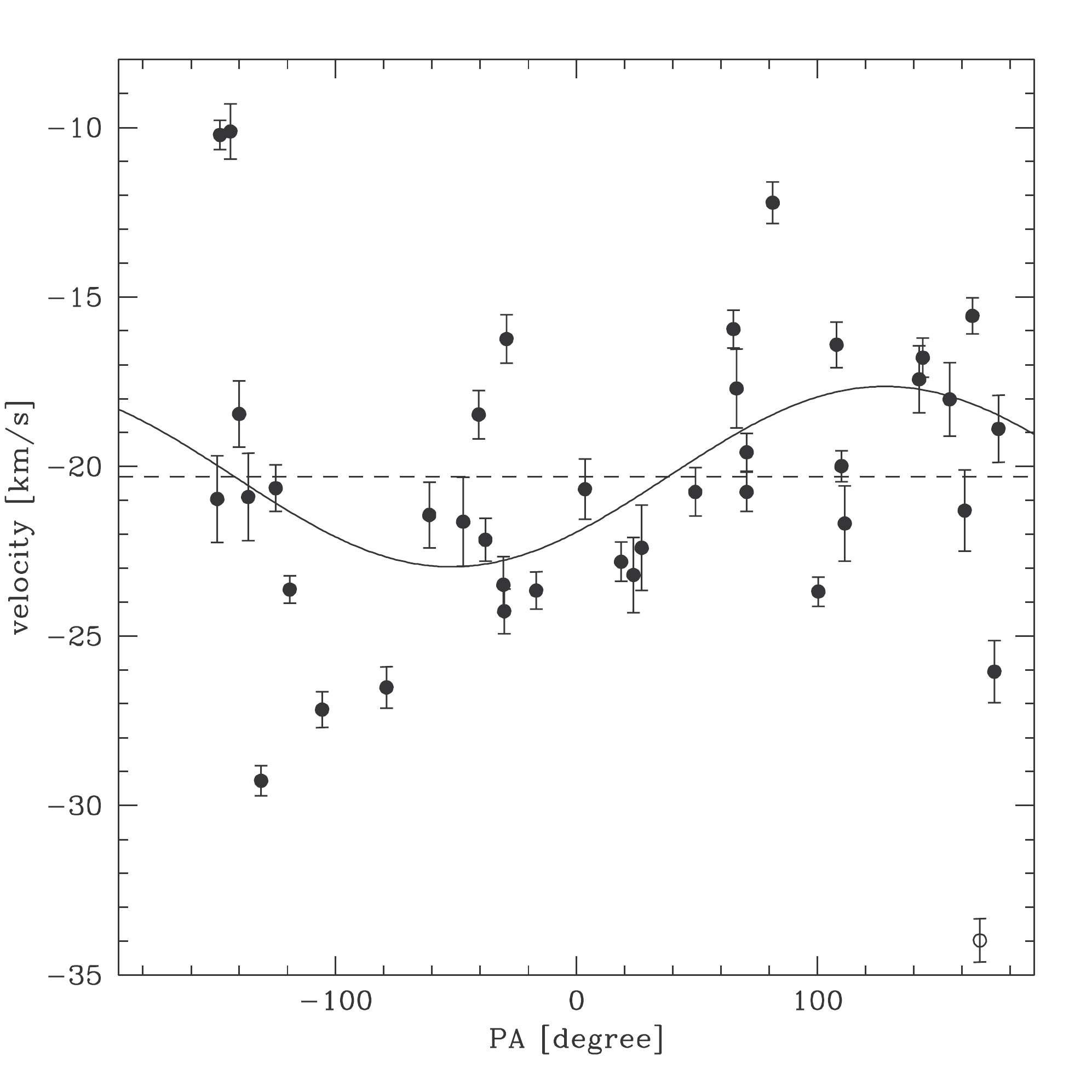}
\end{center}
\caption{Radial velocities as a function of position angle (PA) measured from north (PA=$0^\circ$) towards east. The solid curve shows
our best-fitting rotation curve for NGC~2419. It has an amplitude of $3.26 \pm 0.85$ km/sec, implying significant rotation of the
cluster. The direction of rotation, PA=$40.9^\circ \pm 17.8^\circ$ is, within the errobars $90^\circ$ away from the semi-major axis found in the surface 
density profile by \citet{b07}, PA=$+105^\circ \pm 28^\circ$, supporting the fact that NGC~2419 is rotating.}
\label{fig:rotvel}
\end{figure}

\subsection{Mass modeling and mass-to-light ratio}
\label{sec:res-mas}

The global mass-to-light ratio is calculated from the velocity dispersion and the total cluster luminosity under the assumption that mass 
follows light.
Unless there is primordial mass segregation, this is probably a valid assumption since the relaxation time of NGC~2419 is larger than 
a Hubble time: i.e., the central and half-mass relaxation times are $T_{r0} \approx 10.5$~Gyr and $T_{rh} \approx 19$~Gyr, 
respectively (Djorgovski 1993). In addition, \citet{detal08} found no evidence for mass segregation among the blue straggler stars in NGC~2419, which
supports the assumption that mass follows light.  We also neglect the cluster rotation and assume that the cluster has an isotropic
velocity dispersion. In such a case, we can use the method described in \citet{hetal07} to calculate the mass-to-light ratio.
Their method first deprojects the surface density profile (here a cored S\'ersic profile) to obtain the three-dimensional spatial density distribution, $\rho(r)$. The
spatial density is then used to calculate the potential, $\Phi(r)$, of NGC~2419 and from $\rho$ and $\Phi$, the distribution function, $f(E)$, is 
calculated by assuming spherical symmetry and using eq.\ 4-140a of \citet{bt87}. This distribution function is then used to create
an $N$-body representation (i.e., particle positions and velocities) of NGC 2419 with a given first guess of $M'$ for the cluster mass. We then 
draw a number of $N'$ stars located at the same projected radius of each observed star from this $N$-body model and measure the 
projected velocity dispersion, 
$\sigma_{Mod}$, of the sample stars. The total mass of NGC 2419 can then be calculated by comparing this velocity dispersion with
the observed velocity dispersion according to:
\begin{equation}
 M_C = M' \cdot \sigma^2_{Mod}/\sigma_c^2 \;\; .
\end{equation}
This method has the advantage that it can work for any given observed surface density profile and for any given radial 
distribution of stars. This sets it apart from parameterized formulas which work only for certain density distributions and
usually need the core or global velocity dispersion to calculate the cluster mass. If the stars with radial velocity measurements
are neither concentrated in the cluster core, nor spread out over the cluster in the same way as the cluster stars, using such 
formulae can lead to a bias in the derived cluster mass.

We calculate an $N$-body model with $N=5 \cdot 10^5$ stars in total and extract for each observed star $N'=200$ stars
from this model. Calculating the velocity dispersion for the theoretical model and comparing it with NGC~2419, we 
find a total mass of $M_C = (9.02 \pm 2.22) \cdot 10^5$ M$_\odot$ for NGC 2419 and a mass-to-light ratio of $M/L_V=2.05 \pm 0.50$ in
solar units, using all stars except star S22. If we were to include star S22, the total mass would rise to
 $M_C=(1.09 \pm 0.26) \cdot 10^6$ M$_\odot$  and the mass-to-light ratio would be $M/L_V=2.48 \pm 0.60$ in solar units.
These values are significantly higher than the M/L ratio derived by Olszewski et al. (1993), $M/L_V \approx 0.7 \pm 0.4$~$M_{\odot} / L_{V,\odot}$.
Since our method gives a very similar M/L than what they found if we use their stars as input, the main reason for the discrepant M/L values
seems to be the different input stars.

We finally tested our modeling method by randomly drawing 40 stars from the $N$-body model and then measuring their velocity dispersion 
and fitting the model against them in the same way as was done for the NGC~2419 data. Doing this 100 times, each time with a 
different random realisation of stars, we obtained an average ratio of derived mass-to-light ratio to true mass-to-light ratio of $1.00 \pm 0.01$,
i.e. our modeling method gives an unbiased estimate of the true mass-to-light ratio.

\begin{figure}
\begin{center}
\includegraphics[width=8.5cm]{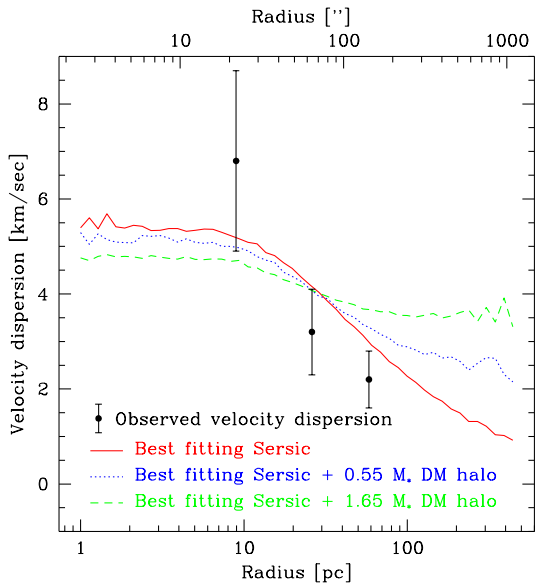}
\end{center}
\caption{Observed velocity dispersion as a function of radius. The solid line shows our prediction based on the best-fitting cored
S\'ersic model. Dotted and dashed lines show the predicted velocity dispersion if we add NFW halos with a scale radius of $R_S=500$ pc 
and masses of $M_{DM}=4 \cdot 10^6$ M$_\odot$ and $M_{DM}=10^7$ M$_\odot$ inside $R_S$ to this model. Models with additional dark 
matter halos significantly overpredict the velocity dispersion in the outer parts, showing that NGC~2419 does not possess a 
dSph-like DM halo.}
\label{fig:radvel}
\end{figure}

In order to compare the observed mass-to-light ratio with the predictions of population synthesis models, we need to know the 
cluster age and metallicity.
Based on an analysis of a high-resolution Echelle spectrum for a single RGB star (10), Shetrone, C\^ot\'e \& Sargent (2001) measured
a metallicity of [Fe/H]=-2.32 for NGC~2419, which is one of the lowest abundances among all Galactic
globular clusters. \citet{sw02} have used horizontal branch and turn off star brightnesses to determine ages of Galactic
globular clusters and derived an absolute age 
of 12.3 Gyr or 12.8 Gyr for NGC~2419, depending on the adopted metallicity scale. For this metallicity and age, the SSP models of 
\citet{m05} predict a mass-to-light ratio of 1.95 for a \citet{k01} IMF, while the \citet{bc03} models predict $M/L_V=1.85$ 
for a \citet{c03} IMF. These values agree remarkably well with the derived mass-to-light ratio. The $M/L_V$ ratio of NGC~2419 is therefore 
entirely compatible with the assumption that the 
cluster formed with a standard stellar mass function. This sets NGC~2419 apart from some UCDs, which have only slightly larger sizes and 
masses, and for which standard mass functions seem to underpredict their mass-to-light ratios \citep{hetal05,metal08}.

The ``normal" mass-to-light ratio of NGC~2419 also argues against the presence of a significant amount of dark matter in the cluster.
In order to test if a possible DM halo could exist at least in the outer cluster parts, we have divided the sample into 
three equally large groups according to their radial distance and calculated 
velocity dispersions separately for each group. Fig.~\ref{fig:radvel} compares velocity dispersions at different radii with the predicted velocity 
dispersion profile from our $N$-body model. It can be seen that the velocity dispersion of the cored S\'ersic model is in reasonable agreement 
with the data. If at all, additional mass is required only in the inner parts
of the cluster, where the predicted velocity dispersion is smaller than the observed one by
about 1$\sigma$, not in the outer parts as might be expected for a dark matter halo. 

We also added different DM halos to our $N$-body model in order to see the effect of these halos on the velocity
dispersion. The DM halos follow NFW profiles with scale radii $R_S=500$ pc,
which agrees with measured scale radii of Galactic dSph galaxies which are generally 
larger than a few hundred pc \citep{wu07}, and had various ratios of dark halo mass inside $R_S$ to the stellar mass of NGC~2419. 
We then calculated a new velocity dispersion
profile for NGC~2419 and fitted it to the data in the same way as described above. From our fit procedure, we derive a best-fitting (stellar) 
M/L ratio and the total stellar and dark mass of NGC~2419.
As can be seen in Fig.~\ref{fig:radvel}, significant amounts of dark matter 
are basically ruled out by our observations since they overpredict the velocity dispersion in the outer cluster parts, 
at least for models with isotropic stellar velocity dispersions. A DM halo with total mass of $M_{DM}=1.65 \cdot M_{Stars}$
overpredicts the velocity dispersion by more than 2$\sigma$ at a radius of R=140$^{\prime\prime}$. This model also has a best-fitting
stellar M/L ratio of only $M/L=1.40 \pm 0.34$, which is 2$\sigma$ below the predictions of the stellar evolution models.
Our observations therefore rule out a DM halo
of more than $M \approx 1 \cdot 10^7$ M$_\odot$ inside 500 pc around NGC~2419, corresponding to a density of 0.02 $M_\odot/pc^3$.
This is significantly lower than the dark matter content of Galactic dSph galaxies inside a similar radius as derived by 
\citet{wetal07} and \citet{setal08}. It is also a factor 5 lower than central dark matter densities of dwarf galaxies as
determined by \citet{getal07}. Our observations therefore argue strongly against the formation of NGC~2419 within a sizeable dark 
matter halo. More radial velocities would help to strengthen our conclusions and test the validity of several assumptions that
we made for our data analysis, in particular that the stellar orbits are isotropic and that the cluster is not mass segregated.

Finally, we note that the low observed velocity dispersion in the outer parts of  NGC~2419 might also be a problem for {\it MOND}. 
According to {\it MOND}, the 
orbits of stars show a difference with respect to standard Newtonian behavior once their acceleration falls below a critical acceleration
$a_0 \sim 1.2 \cdot 10^{-8}$ cm/sec$^2$ \citep{sm02}. While the acceleration of stars near the half-mass radius of NGC~2419 is still
higher than $a_0$, it falls significantly below $a_0$ in the outer parts of NGC~2419. At the radius of our outermost datapoint in 
Fig.~\ref{fig:radvel}, the internal acceleration is only one third of $a_0$, while the external acceleration is only one tenth
of $a_0$, so one would expect that the observed velocities start to deviate 
significantly from the prediction of our $N$-body model which was created based on Newtonian dynamics. 
This is not
seen in our data. In fact, as shown by Milgrom (1994), if one neglects the external 
field of the Milky Way, {\it MOND} predicts that the line-of-sight velocity dispersion should level off at a value 
$\sigma_{min}=0.471 \sqrt{G M_C/a_0}$ at large distances. If we use the above calculated cluster mass, we 
find $\sigma_{min}=2.6$ km/sec  for NGC~2419. Our outermost datapoint is already 
probing this value, so acquiring additional data at even larger radii might prove a powerful 
way to test the validity of {\it MOND}. 

\section{Conclusions}
\label{sec:concl}

We have measured precise radial velocities for $\approx$ 40 stars in the outer halo globular cluster NGC 2419 using the Keck telescope
High Resolution Echelle Spectrometer. We find a mean cluster velocity of $<\!\!v_r\!\!>=-20.3 \pm 0.7$ km/sec and an internal 
velocity dispersion of $\sigma = 4.14 \pm 0.48$ km/sec. Our observations also reveal a slight cluster
rotation with amplitude $3.26 \pm 0.85$ km/sec around a position angle of $40.9 \pm 17.8$ degrees. From a comparison of the 
measured velocity dispersion to the velocity dispersion of a
spherically symmetric model for NGC~2419, we find a total mass of $M_C = (9.01 \pm 2.22) \cdot 10^5$ M$_\odot$ and a mass-to-light ratio 
of $M/L_V = 2.05 \pm 0.50$ in solar units. This value is entirely compatible with the one expected for a stellar system at the metallicity of 
NGC~2419 following a standard mass function like \citet{k01} or \citet{c03}.  

NGC 2419 therefore does not show any dynamical evidence of a significant depletion of low-mass stars, which is consistent with expectations given the large relaxation and dissolution time for this cluster. Similar M/L ratios which are in agreement with standard
mass functions have been found for other galactic globular clusters with large relaxation and dissolution times like Omega 
Cen ($M/L_V=2.5$ \citet{vdv06}). This sets Milky Way globular clusters apart from 
UCDs for which standard mass functions on average seem to underpredict mass-to-light ratios \citep{metal08}.
If real, this points to a significant change in the star formation process
occurring at a characteristic mass of $M_C \approx 2 \cdot 10^6$ M$_\odot$ (e.g., Ha{\c se}egan et al. 2005; Mieske et al. 2008,
Dabringhausen et al.\ 2009).
Such a transition might arise, for example, if more massive clusters become
optically thick to far infrared radiation and are born with top heavy initial mass functions \citep{m08}.

We do not find any evidence for the presence of substantial amounts of dark matter in NGC~2419. Since 
both the depletion of dark matter from the cluster center due to two-body relaxation and dynamical friction \citep{bm08}, 
as well as tidal stripping 
of dark matter from the cluster halo are unlikely, this indicates that NGC~2419 did not contain substantial amounts of dark matter at the time of formation. 
NGC~2419 is, however, one of the most likely candidates for a globular cluster to have formed with an associated 
dark matter halo given to its quite low metallicity. 
Our non-detection of dark matter in this cluster therefore supports the hypothesis that globular clusters, as a rule, did not form at the
centers of dark matter halos. Instead, an origin driven by gas dynamical processes during mergers between galaxies \citep{betal08} or
proto-galactic fragments seems to be the more likely explanation for the formation of even the lowest
metallicity globular clusters. 

We also find a slightly larger than expected velocity dispersion in the central regions of the
cluster, which might point to a significant binary fraction or additional unseen matter in the core. At the same time, the rather 
low velocity dispersion in the outer regions of the cluster could pose a problem for alternative gravitational theories
like {\it MOND}. These conclusions rely on several assumptions that we had to make for our modeling, in particular that 
stellar orbits are isotropic throughout the cluster and that the stellar mass-to-light ratio is constant with radius. Acquiring additional radial velocities
would help test the validity of these assumptions and strengthen our conclusions.

\section*{Acknowledgements}

We thank Tad Pryor and Ed Olszewski for providing their finding charts for NGC~2419. We also thank Iskren Georgiev and an anonymous referee
for comments which improved the presentation of the paper. This study was based on observations obtained at the W. M. Keck 
Observatory, which is operated jointly by the California Institute of Technology and the University of California. We are 
grateful to the W. M. Keck Foundation for their vision and generosity. SGD acknowledges a partial support
from the NSF grant AST-0407448, and the Ajax Foundation.

\label{lastpage}

\end{document}